\documentclass[aps,prl,reprint,superscriptaddress]{revtex4-1}
\usepackage{hyperref}
\usepackage{graphicx}
\usepackage{enumerate}
\hypersetup{
    pdfnewwindow=true,      
    colorlinks=true,       
    linkcolor=blue,          
    citecolor=blue,        
    filecolor=blue,      
    urlcolor=blue           
}

\usepackage{amsmath}
\usepackage{amssymb}
\usepackage{amsfonts}

\usepackage{color}

\def\be{\begin{equation}}
\def\ee{\end{equation}}

\def\lsim{~\rlap{$<$}{\lower 1.0ex\hbox{$\sim$}}}
\def\gsim{~\rlap{$>$}{\lower 1.0ex\hbox{$\sim$}}}

\providecommand{\aap}[0]{A\&A~}

\providecommand{\mnras}[0]{MNRAS~}

\providecommand{\nat}[0]{Nature~}

\providecommand{\prd}{Phys. Rev. D.~}

\providecommand{\nar}[0]{New Astron. Rev.~}

\begin{document}

\title{G2 can Illuminate the Black Hole Population near the Galactic Center}

\author{Imre Bartos}
\email[]{ibartos@phys.columbia.edu}
\altaffiliation{Columbia Science Fellow}
\affiliation{Department of Physics, Columbia University, New York, NY 10027, USA}
\affiliation{Columbia Astrophysics Laboratory, Columbia University, New York, NY 10027, USA}
\author{Zolt\'an Haiman}
\affiliation{Columbia Astrophysics Laboratory, Columbia University, New York, NY 10027, USA}
\affiliation{Department of Astronomy, Columbia University, New York, NY 10027, USA}
\author{Bence Kocsis}
\affiliation{Harvard-Smithsonian Center for Astrophysics, Cambridge, MA 02138, USA}
\author{Szabolcs M\'arka}
\affiliation{Department of Physics, Columbia University, New York, NY 10027, USA}
\affiliation{Columbia Astrophysics Laboratory, Columbia University, New York, NY 10027, USA}

\begin{abstract}
Galactic nuclei are expected to be densely populated with stellar and intermediate mass black holes. Exploring this population will have important consequences for the observation prospects of gravitational waves as well as understanding galactic evolution. The gas cloud G2 currently approaching Sgr A* provides an unprecedented opportunity to probe the black hole and neutron star population of the Galactic nucleus. We examine the possibility of a G2-black hole encounter and its detectability with current X-ray satellites, such as Chandra and NuSTAR. We find that multiple encounters are likely to occur close to the pericenter, which may be detectable upon favorable circumstances. This opportunity provides an additional, important science case for leading X-ray observatories to closely follow G2 on its way to the nucleus.
\end{abstract}

\maketitle

\noindent
{ \bf 1. Introduction} --- Stellar remnants are expected to migrate towards supermassive black holes (SMBHs) in galactic nuclei due to dynamical friction \cite{1993ApJ...408..496M}. In the case of the Milky Way, a steep density cusp of $\sim20,000$ black holes (BHs) and a similar number of neutron stars (NSs) is predicted to be present in the central parsec \cite{1993ApJ...408..496M,2000ApJ...545..847M,2006ApJ...649...91F,2009MNRAS.395.2127O}. Much heavier, so called intermediate mass-BHs (IMBHs; $M\gg10\,$M$_\odot$), may also be present in the galactic nucleus due to mass segregation with stars (e.g., \cite{2012ApJ...752...67K}).

Black holes play an important role in a wide range of fields. They are important in galaxy formation, accretion and high-energy astrophysics, stellar evolution, gravitational-wave astrophysics, as well as general relativity. Nevertheless, despite their expected high densities in galactic nuclei, their discovery remains elusive.

The density of compact objects in galactic centers has important implications for
the detection of gravitational waves (GWs). The close encounter of compact objects can
result in gravitational radiation detectable by Earth-based interferometers
(e.g., \cite{2006ApJ...648..411K}). Highly eccentric binaries may form due to
GW losses in close encounters, or due to the secular
gravitational effects of the central supermassive black hole \cite{2012arXiv1206.4316N,2012ApJ...757...27A}.
The rate of such sources can be sufficiently high for many detections
with advanced GW detectors \cite{2009MNRAS.395.2127O}.
Since the GW signatures of these distinct sources differ significantly
from traditional compact binary mergers \cite{2012arXiv1212.0837E,2012PhRvD..85l3005K},
their detection will require the development of specialized GW search algorithms.

G2, a dense gas cloud with an estimated mass of $3M_\oplus$ is approaching the Galactic center on a highly eccentric trajectory \cite{2012Natur.481...51G,2012ApJ...750...58B,2013ApJ...763...78G}. It will reach its pericenter of $\sim10^{-3}\,$pc around September 2013 \cite{2013ApJ...763...78G}. Through its plausible encounter with BHs in the galactic nucleus, G2 provides a unique opportunity to probe this BH population that is difficult to observe otherwise.

In this paper we examine the detectability of a BH-G2 encounter in the Galactic center. We show that multiple such encounters can occur near the pericenter, and we estimate their luminosity, and discuss their detectability.

\vspace{3 mm}
\noindent
{\bf 2. Probability of encounter. ---}
Given the large expected population of stellar-mass BHs in the vicinity of Sgr A*, G2 may encounter some of these BHs which can be detectable via its X-ray emission. To estimate the probability of such encounters, we adopt the expected BH population derived by O'Leary et al. (\cite{2009MNRAS.395.2127O}; Model B). In their numerical model, O'Leary et al. find that BHs segregate and form a steep density cusp around Sgr A*, reaching BH number densities $n_{\rm bh}>10^{9}\,$pc$^{-3}$ within the central $10^{-3}$\,pc. Depending on the mass distribution of the more massive BHs, the density cusp may become even steeper \cite{2009ApJ...698L..64K}. The cusp is dominated by the most massive BHs in the inner region.

\begin{figure}
\begin{center}
\resizebox{0.475\textwidth}{!}{\includegraphics{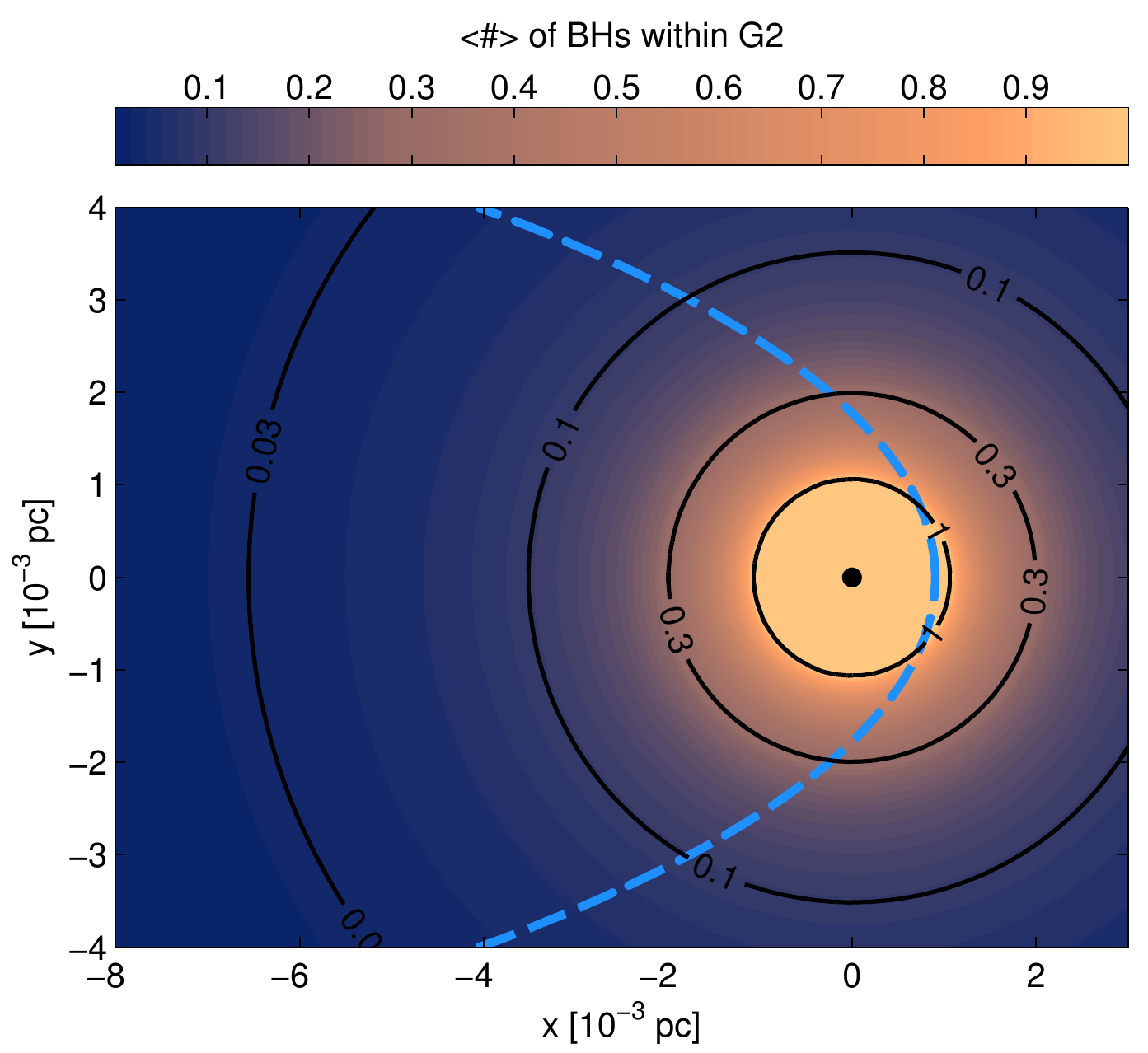}}
\end{center}
\caption{Expected number of stellar-mass BHs within G2 (with volume $2\times10^{-10}$\,pc$^{3}$) as a function of the location of G2 ([0,0] corresponds to Sgr A*). The dashed line indicates the projected trajectory of G2.}
\label{fig:BHinG2density}
\end{figure}

First, we calculate the probability that a BH is present in the nominal volume $V_{\rm G2}\sim10^{-9}\,$pc$^{3}$ of G2 (based on G2's radius $r_{\rm G2}\sim6\times10^{-4}\,$pc as of early 2012 \cite{2012ApJ...750...58B}), as a function of the position of the cloud. We adopt the orbital parameters of the cloud from Gillessen et al. \cite{2013ApJ...763...78G}. We evaluate $n_{\rm bh}$ at positions along this orbit, and conservatively assume that it is constant inside the entire volume of G2.  We also assume that the volume and density of the cloud remains unchanged along its orbit (the density could significantly increase due to ram pressure and/or thermal pressure from the surrounding hot medium; nevertheless, it is estimated to remain constant due to tidal effects countering this increase; see \cite{2012ApJ...750...58B}). The results, shown in Figure \ref{fig:BHinG2density}, indicate that close to its pericenter, G2 will host, on average, more than one BHs.

Next, we calculate the expected total number $\langle N_{\rm bh}\rangle$ of BHs encountered by G2 along its parabolic trajectory around Sgr A*. We take an elongated cloud as described below. For simplicity we assume that BHs enter the cloud perpendicular to its long side (which is the limiting case for eccentricity $e_{\rm G2} \rightarrow 1$), corresponding to a projected surface area of $A_{\rm G2}^{\rm side}\sim10^{-6}-10^{-5}$\,pc$^2$. We integrate the BH encounter rate for a full parabolic orbit of G2 (lasting $\approx200$ years) labeled  $S$, with respect to time to obtain
\begin{equation}
\langle N_{\rm bh}\rangle \approx \oint_S\,n_{\rm bh}(r)\,A_{\rm G2}^{\rm side}v_{\rm bh}(r)\,\mbox{d}t\sim 10,
\end{equation}
where $r=r(t)$ is the distance between G2's center and Sgr A*, $v_{\rm bh}(r)$ is the Keplerian
velocity of a BH on a circular orbit at distance $r$ from Sgr A*, and $\mbox{d}t$ is the orbital time of G2 between $r$ and $r+\mbox{d}r$. The majority of the encounters occur close to the pericenter ($\sim16$ of them is expected within $\pm0.5$\,yr from the pericenter).

It is also interesting to estimate the characteristic duration of a G2-BH encounter. For this, we take into account the fact that as it moves towards the pericenter, the cloud is stretched by tidal effects \cite{2012ApJ...750...58B}. We adopt an oblate ellipsoidal shape for the cloud with aspect ratio 1:10 (i.e. eccentricity of $\sim 0.995$), with the long axis oriented parallel to its trajectory (see Fig. 4 of \cite{2012ApJ...750...58B}). Given its elongated shape and the highly eccentric trajectory, BHs on circular orbit around Sgr A* most likely cross G2 through its smaller cross section, i.e. perpendicular to G2's trajectory. Given the Keplerian velocity $v_{\rm k}\approx3800$\,km\,s$^{-1}$ of a body around Sgr A* at the pericenter distance of G2 and considering the ellipsoidal shape of the cloud, the typical crossing time for BHs through G2 is $t_{\times}\approx\,$month. This is long enough to be captured with short exposures spaced $\lsim$ month apart, and to be followed up in longer exposures over a $\sim$month.

IMBHs can be born through multiple channels. The collapse of pregalactic (PopIII) stars can lead to the formation of IMBHs. Such process can add $\sim50(M_{\rm imbh}/150\,$M$_{\odot})$ IMBHs in the vicinity of the Galactic nucleus \cite{2001ApJ...551L..27M}. Runaway growth in dense star clusters can also produce IMBHs \cite{2002ApJ...576..899P,2006MNRAS.368..141F}. This process can result in $\sim50$ IMBHs of masses $\sim10^3\,$M$_\odot$ within $\sim10\,$pc of the Galactic nucleus. IMBHs can also be formed via the merger of stellar-mass BHs, whose masses can further grow, e.g., via stellar collisions \cite{2006ApJ...637..937O}. After formation through these distinct channels, IMBHs move towards the Galactic nucleus due to mass segregation (e.g., \cite{2012ApJ...752...67K}). IMBHs could also be born and migrate inward in the SMBH accretion disk \cite{Levin,Barry}.

\vspace{3 mm}
\noindent
{\bf 3. Expected Luminosity \& Spectrum. ---}
Here, we estimate the expected luminosity of a BH with mass $M_{\rm bh}$ moving through G2 with a relative velocity $\Delta v$, mostly following the discussion of \cite{2007MNRAS.377.1647N}. For the rate of gas accretion by the BH, we adopt the Bondi-Hoyle-Lyttleton (BHL) rate (e.g., \cite{2004NewAR..48..843E}). Given the low temperature of the gas [$\mathcal{O}(10^3\,\mbox{K})$], the gas sound speed is negligible compared to the expected relative velocities [$\mathcal{O}(10^3\,\mbox{km/s})$], and the BHL accretion rate can be written as
\begin{equation}
\dot{M}_{\rm BHL} \sim 4\pi \rho_{\rm c} \, G^2 \, M_{\rm bh}^2 \, \Delta v^{-3},
\label{equation:BondiAccretion}
\end{equation}
where $\rho_{\rm c}$ is the gas density. For representative values $\rho_{\rm c}\sim10^{-18}\,$g\,cm$^{-3}$, $M_{\rm bh}\sim10$M$_\odot$ and $\Delta v\sim10^3\,$km\,s$^{-1}$ \cite{2012ApJ...750...58B}, the accretion rate will be $\dot{M}_{\rm BHL}\sim10^{-6}\dot{M}_{\rm Edd}$, where $\dot{M}_{\rm Edd}\approx7\times10^{-15}(M_{\rm bh}/10\,\mbox{M}_{\odot})\,\mbox{M}_{\odot}\,$s$^{-1}$ is the fiducial rate corresponding to the Eddington luminosity. Note that the Bondi radius
$R_{\rm B}\approx 3\times 10^{11} (M/10{\rm M_\odot}) (\Delta v/10^3{\rm km s^{-1}})^{-2}$\,cm
is much smaller than the size of G2 $R_{\rm G2}\approx 2\times10^{15}cm$, validating the above accretion rate.

For low accretion rates, $\dot{M}_{\rm BHL} \lesssim 0.01 \dot{M}_{\rm Edd}$, only a small fraction of the accreted material is converted into electromagnetic radiation, while most of the energy is lost at the event horizon or through winds. We take the X-ray radiative efficiency of the accretion to be $\epsilon_{\rm x}\approx\dot{M}_{\rm BHL}/\dot{M}_{\rm Edd}$ (assuming that $\sim10\%$ of the bolometric luminosity is in the X-ray band) \cite{1995ApJ...452..710N,2007MNRAS.377.1647N,2008NewAR..51..733N}. The expected X-ray luminosity $L_{\rm x} = \epsilon_{\rm x}\,\dot{M}_{\rm BHL}c^2$ of the BH within G2 is thus
\begin{equation}
L_{\rm x} \approx 3\times10^{31}\,\mbox{erg\,s}^{-1}\,\,\rho_{18}^2\left(\frac{M_{\rm bh}}{100\,\mbox{M}_{\odot}}\right)^3\left(\frac{\Delta v}{10^3\,\mbox{km\,s}^{-1}}\right)^{-6},
\label{equation:Lx}
\end{equation}
where $\rho_{18}=\rho_c/(10^{-18}\,$g\,cm$^{-3})$ (the estimated cloud density is $\rho_{18}\approx1$ \cite{2012ApJ...750...58B}).
We note that there are alternatives to the accretion/radiation model discussed above. In general, accretion onto a BH may strongly depend on the topology of magnetic fields (e.g., \cite{2012MNRAS.423.3083M,2012MNRAS.426.3241N}) or the angular momentum of the accreted gas \cite{2013ApJ...767..105L}. The BH spin can also qualitatively affect accretion, potentially giving rise to powerful jets (e.g., \cite{2012MNRAS.426.3241N}). Alternative models could predict either a much higher or much lower luminosity (see, e.g., \cite{2011ApJ...736L..23W,2013ApJ...767..105L,SS73}).

We showed above that the majority of the encounters is expected to occur in the inner region of G2's trajectory, where its speed approaches $\sim6300\,$km\,s$^{-1}$ \cite{2013ApJ...763...78G}. Since the expected luminosity greatly depends on $\Delta v$, those BH encounters for which the BH co-moves with G2 at its pericenter (where G2's direction is tangential) will result in significantly increased luminosity. At the pericenter $r_{\rm peri}\approx10^{-3}\,$pc of G2, the speed of a BH in a circular orbit is $v_{\rm peri}^{\rm bh}\approx 4500\,$km\,s$^{-1}$, therefore a BH whose orbital plane is close to the orbital plane of G2 can yield $\Delta v\sim10^3\,$km\,s$^{-1}$. Furthermore, if some BHs are on highly elliptical orbits similar to G2's, they can encounter G2 with an even lower $\Delta v$.

We adopt a hard radiation spectrum of $F_{\nu}\propto \nu^{-1}$ (where $\nu$ is the photon frequency), which is consistent with the typical spectrum of black-hole X-ray binaries in their quiescent state (e.g., \cite{1998ApJ...505..854E,2002MNRAS.334..553A}). We consider such an X-ray spectrum to be present in the photon energy range 0.1-100\,keV (c.f. \cite{1998ApJ...505..854E,2002MNRAS.334..553A,1999ApJ...520..298Q,2004ApJ...617L..49K}).

It is useful to compare the above luminosities to the expected luminosity of compact objects due to their interaction with the accretion disk of Sgr A*. Adopting accretion disk properties by Sadowski et al. \cite{2013arXiv1301.3906S}, the accretion disk at G2's pericenter distance from Sgr A* has a maximum density $\rho\sim10^4$\,cm$^{-3}$ and temperature $T\sim6\times10^8$\,K. This temperature corresponds to an effective $\Delta v$ of $4000$\,km\,s$^{-1}$. Given that Sgr A*'s accretion disk is much sparser than G2 at and beyond the pericenter, even a compact object corotating with the accretion disk will be significantly fainter than one that encounters G2.

\vspace{3 mm}
\noindent
{\bf 4. Detectability. ---}
To estimate the detectability of G2-BH and G2-NS encounters, we consider the Chandra X-ray observatory \cite{1999astro.ph.12097W}. Chandra's high directional resolution and sensitive energy band ($0.1-10$\,keV) makes it suitable for such analysis. For observations in the vicinity of the Galactic center, we adopt a detection limit of $10^{32}$\,erg\,s$^{-1}$ ($2-8\,$keV) for an observation time of one week. Note that G2 is already inside the distance corresponding to Chandra's point spread function from the Galactic center; this limit is therefore set by the shot noise from SgrA*'s own X-ray emission (c.f. Chandra's sensitivity in other directions \cite{2003ApJ...589..225M}).

\begin{figure}
\begin{center}
\resizebox{0.475\textwidth}{!}{\includegraphics{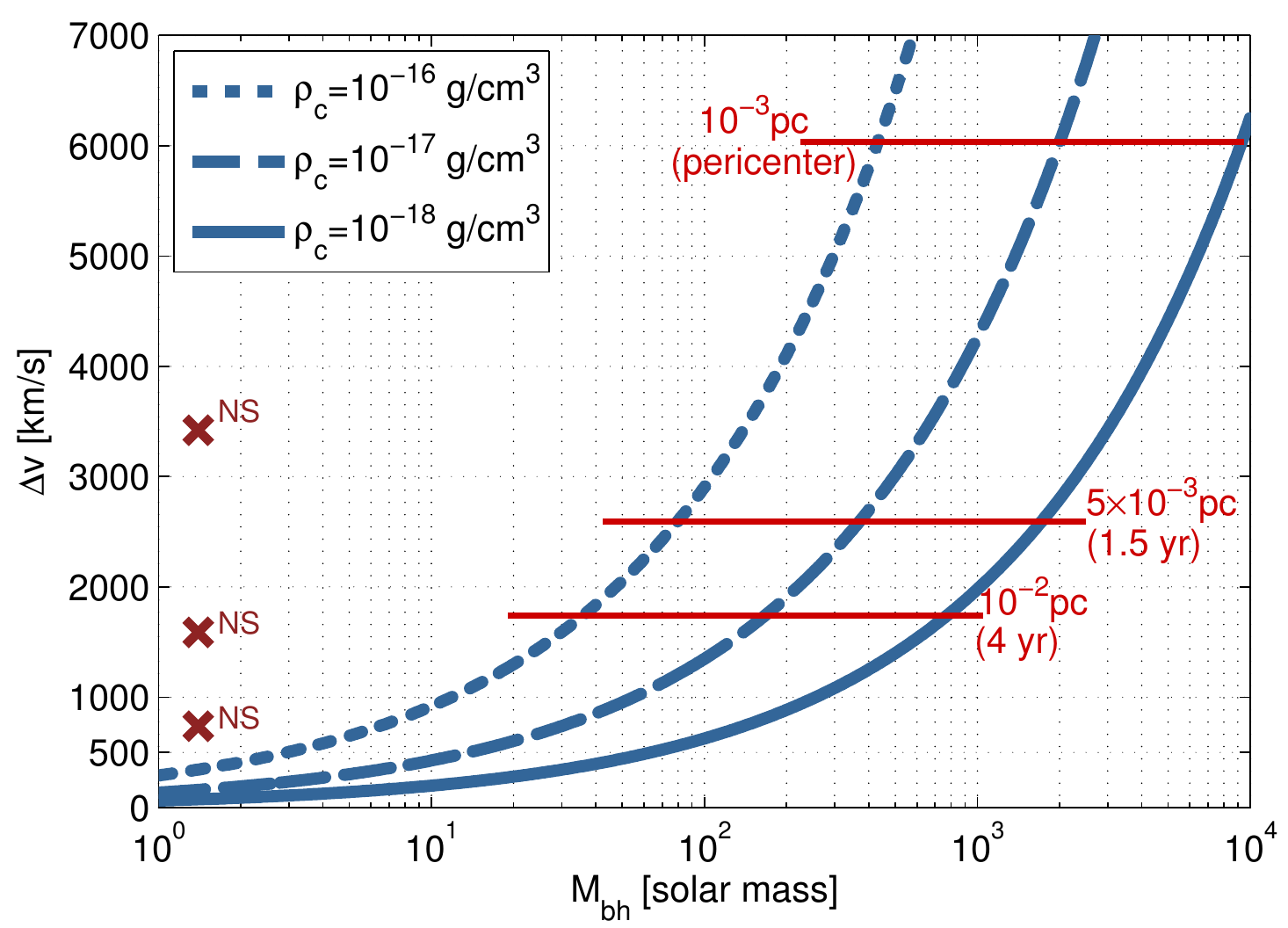}}
\end{center}
\caption{Maximum relative G2-BH velocity ($\Delta v$) at which a G2-BH encounter is detectable with Chandra, as a function of BH mass, for different G2 densities. For reference, the horizontal lines indicate the orbital speeds reached by G2 at different distances from Sgr A* and times from reaching the pericenter (shown next to the lines). G2-NS (1.4\,M$_\odot$) velocities are also indicated with '$x$' for $\rho_{18}=\{1,10\}$.}
\label{fig:mindv}
\end{figure}

To estimate the detectability of a G2-BH encounter, we determine the parameters ($M_{\rm bh}$ \& $\Delta v$) for which the encounter is detectable by Chandra. We take $L_{\rm X}$ from Eq. \ref{equation:Lx}, and assume that an \emph{effective} $20\%$ of the total X-ray radiation is in the $2-8\,$keV band. Our results are shown in Figure \ref{fig:mindv}.

The figure shows that, for constant G2 density ($\rho_{18}=1$), stellar-mass BHs ($M_{\rm bh}\sim10\,$M$_\odot$) need to move along with G2 to be detectable. Adopting a simplified BH-velocity distribution model from \cite{2000ApJ...545..847M} with isotropic directional and constant velocity distribution up to the escape velocity, the expected number of detectable G2-stellar-mass-BH encounter is negligible. For $\gtrsim1$ expected detection rate, a measurement needs to be able to detect such an encounter with $\Delta v \lesssim 2900$\,km\,s$^{-1}$.

The encounter of G2 with an IMBH is more promising. IMBHs on co-planar orbits can be detected at any mass, while for higher masses even IMBHs with less aligned orbits can be detected. A potential increase in G2's density (e.g., at its outer shell \cite{2012Natur.481...51G}) can further mitigate the dependence of detectability on orbital alignment. Using the BH-velocity distribution model described above (see \cite{2000ApJ...545..847M}), one gets $\ge50\%$ detection rate for G2-IMBH encounters for BH masses for which $\Delta v \leq 4400$\,km\,s$^{-1}$ is detectable. With constant G2 density $\rho_{18}=1$, this corresponds to $M_{\rm bh}\gtrsim5\times10^3$\,M$_{\odot}$, i.e. if encounters with BHs above $5\times10^3$\,M$_{\odot}$ occur, they provide a promising target for detection.

The potential observation of G2-NS encounters is another interesting possibility. While the radiative efficiency of accreting BHs is severely reduced due to the energy lost through the event horizon, NSs' surface can reradiate the energy of the infalling matter, making NSs' radiative efficiency significantly higher (e.g., \cite{1999ApJ...520..276M}). We adopt a NS radiative efficiency of $\epsilon_{\rm ns}\sim0.1$
(e.g., \cite{1999ApJ...520..276M,2008NewAR..51..733N}). While the accretion rate of highly-magnetized, rotating NSs can be significantly reduced compared to the BHL rate, the magnetic field strengths of isolated NSs is poorly constrained and can be negligible \cite{2003ApJ...594..936P}. Here we consider such NSs with negligible magnetic fields. Strong magnetic fields could prevent a rotating NS crossing G2 from accreting \cite{1975A&A....39..185I}. The X-ray luminosity of such a NS with mass $M_{\rm ns}=1.4\,$M$_\odot$ accreting at the BHL rate (Eq. \ref{equation:BondiAccretion}) will be
\begin{equation}
L_{\rm X}^{\rm ns} \approx 4\times10^{31}\,\mbox{erg\,s}^{-1}\,\,\rho_{18}\left(\frac{\Delta v}{10^3\,\mbox{km\,s}^{-1}}\right)^{-3}.
\end{equation}
If, similarly to the case of BHs, $\sim20\%$ of this radiation is in the $2-8\,$keV range, then G2-NS encounters are detectable with Chandra for $\Delta v \lesssim 2000\,$km\,s$^{-1}$.

To identify a G2 encounter in the background, one can utilize the expected duration ($\gtrsim1$\,month) of such an encounter to exclude other, permanent X-ray point sources. X-ray flares due to Sgr A*'s accretion disk can also be excluded, given their typical duration of $\ll1\,$month.

We note that the detectability of encounters could be reduced by the X-ray emission from interaction of G2 and Sgr A*'s accretion disk, which could reach or exceed the quiescent X-ray luminosity of Sgr A* close to G2's pericenter \cite{2012Natur.481...51G,2013ApJ...763...78G}. Nevertheless, the luminosity of this emission is uncertain, and it greatly depends on, e.g., the unknown density of the hot gas along the trajectory of G2. Further, this X-ray emission is expected to slowly change with time compared to G2-BH encounters.

\vspace{3 mm}
\noindent
{\bf 5. Conclusion. ---}
G2 represents a unique opportunity to probe the compact object
population of the Galactic nucleus. We find that $\mathcal{O}(10)$ of
encounters of G2 with stellar mass BHs can occur in the vicinity of
G2's pericenter ($\sim\pm 6\,$months). Some of these encounters can be
observable with the Chandra X-ray telescope.  The point-source
sensitivity of NuSTAR \cite{2009SPIE.7437E..10K} in the 1-10keV range is $\mathcal{O}(10^3)$
lower than Chandra's.  However, if the encounter produces emission
much harder than the $\nu^{-1}$ spectrum adopted here, NuSTAR could
detect at least the most extreme configurations. 
Since an encounter lasts typically more than a month, multiple
encounters can occur simultaneously. Furthermore, encounters with
either NSs or with intermediate-mass BHs ($M_{\rm
  bh}\gsim{\rm few}\times 100~{\rm M_\odot}$) may also occur and may be detectable.

For a constant cloud density of $10^{-18}\,$g\,cm$^{-3}$, the
encounter with a stellar-mass BH ($\sim10$\,M$_\odot$) may be
detectable only if the relative velocity of G2 and the BH is $\Delta
v\lsim 250~{\rm km~s^{-1}}$, significantly below the characteristic
Keplerian speeds. In practice, this can occur if the BH is on
elliptical orbit that is fortuitously aligned with the orbit of G2. Encounters with IMBHs have a better chance of being detected. The requirement on the IMBH's and G2's relative velocity gradually loosens
for higher BH masses. Gor M$_{\rm bh}\gtrsim5000\,$M$_\odot$,
$50\%$ of the encounters become detectable (where we adopted a BH-velocity distribution model from \cite{2000ApJ...545..847M}).
NSs colliding with G2 with $\Delta v
\lesssim750\,$km\,s$^{-1}$ may also be detectable.

The existence of stellar-mass BH in galactic nuclei depends on the
details of stellar populations and stellar dynamics in the bulge.
Likewise, the existence of more massive IMBHs in the bulge, as well as
their abundance and characteristic masses, depend on models of PopIII
star formation in the early universe, and on models of subsequent
hierarchical galaxy assembly. Finally, there are large uncertainties
in the expected luminosity and spectrum arising from low-angular
momentum accretion in a high-velocity encounter. A potential detection
would therefore be a probe of both galaxy formation and of BH
accretion models.

The authors thank Akos Bogdan, Charles Hailey, Jules Halpern, Avi Loeb, Brian Metzger, Jeremiah Ostriker and Frits Paerels for helpful discussions. The Columbia Experimental Gravity group is grateful for the generous support from Columbia University in the City of New York and from the National Science Foundation under cooperative agreement PHY-0847182. ZH acknowledges support from NASA grant NNX11AE05G.

\bibliographystyle{h-physrev}

\end{document}